\documentstyle[12pt,aasms4]{article} 
\def\etal{{et~al.}\ }

\def\vol#1  {{{#1}{\rm,}\ }}

\def\etal{et al.\ }

\def\clock{\count0=\time \divide\count0 by 60
     \count1=\count0 \multiply\count1 by -60 \advance\count1 by \time
     \number\count0:\ifnum\count1<10{0\number\count1}\else\number\count1\fi}

\begin{document}
\title{\bf A Critical Test of Topological Defect Models: Spatial Clustering of Clusters of Galaxies}
\author{Renyue Cen}
\vskip 0.5cm
\centerline{Princeton University Observatory, Princeton, NJ 08544}
\vskip 0.5cm
\centerline{cen@astro.princeton.edu}

\begin{abstract}
Gaussian cosmological models, typified by the inflationary cold dark
matter models,
and non-Gaussian 
topological defect based cosmological models,
such as the texture seeded model,
differ in the origin of large-scale cosmic structures.
In the former 
it is believed that peaks at appropriate scales 
in the initial high density field 
are the sites onto which matter accretes and collapses to form
the present galaxies and clusters of galaxies,
whereas in the latter these structures
can form around the density perturbation seeds
(which are textures in the texture model).
Textures initially are randomly distributed on scales larger
than their size, in sharp contrast to
the initial high density peaks in the Gaussian models
which are already strongly clustered 
before any gravitational evolution has occured.
One thus expects
that the resultant correlation of large cosmic objects
such as clusters of galaxies in the texture model
should be significantly weaker than its Gaussian counterpart.

We show that an $\Omega_0=1$ biased $b=2$ (as required
by cluster abundance observations)
texture model (or any random seed model) predicts a
two-point correlation length of $\le 6.0h^{-1}$Mpc 
for rich clusters, independent of richness. 
On the other hand,
the observed correlation length for rich clusters
is $\ge 10.0h^{-1}$Mpc at an approximately $2\sigma$ confidence level.
It thus appears that the global texture cosmological model
or any random seed cosmological models 
are ruled out at a very high confidence ($>3\sigma$).
\end{abstract}


\section{Introduction}

Cosmological models seeded by topological defects such as 
cosmic strings (Zel'dovich 1980; Vilenkin 1981,1985),
global textures (Turok 1989) 
and global monopoles (Barriola \& Vilenkin 1989; Bennett \& Rhie 1990)
have been proposed as
an alternative to inflationary models,
under the general gravitational instability paradigm of 
structure formation.
Inflationary models such as the 
cold dark matter (CDM) models are Gaussian
in the sense that the one-point density probability function
is Gaussian distributed,
generated by random quantum fluctuations
in the early universe 
(Guth \& Pi 1982; 
Albrecht \& Steinhardt 1982;
Hawking 1982; 
Linde 1982;
Starobinsky 1982;
Bardeen, Steinhardt \& Turner 1983).
In contrast, topological defect based models, taking
the texture model as an example throughout this paper,
are non-Gaussian and highly skewed.
More importantly, it is thought that, in the Gaussian models,
cosmic entities such as galaxies and clusters of galaxies
form around high peaks in the initial density field,
identifiable after 
smoothing the density field
by a window of the size of the objects of interest.
This structure formation scheme in Gaussian models,
initially put forth by Kaiser (1984) to explain the 
enhanced Abell cluster correlation and named the ``biased" 
structure formation mechanism,
yields a clustering of clusters of galaxies 
substantially stronger than matter or galaxies
(e.g., Peacock \& Heavens 1985; 
Barnes \etal 1985;
Bardeen \etal 1986;
Martinez-Gonzalez \& Sanz 1988;
Frenk \etal 1990).
On the other hand, 
the texture model (Turok 1989)
predicts that textures,
being the sites where galaxies or clusters of galaxies later form,
are uncorrelated and randomly distributed {\it initially} on
scales larger than their horizon size (Spergel \etal 1991),
which in the case of clusters is about $10h^{-1}$Mpc.
In this paper we relate this initial property of textures
to the final spatial 
correlational properties of clusters of galaxies
(Bahcall \& Soneira 1983; Klypin \& Kopylov 1983).
We identify clusters of galaxies as a good tool
because they are likely to be less affected than
galaxies by non-gravitational processes.
Consequently, calculations of their clustering properties
under the gravitational instability scenario are likely
to be valid.
In addition, the scales where the cluster-cluster
correlation function is reliably measured observationally
have not yet left
the linear regime and therefore 
some simple linear theory relations may be employed.
We show that the two-point cluster-cluster
correlation length in the texture model
is in the range $5.0-13.0h^{-1}$Mpc for an unbiased $b=1$ model,
and proportional approximately inversely to $b$.
In comparison, the observed correlation length is $\ge 10h^{-1}$Mpc
(at an approximately $2\sigma$ level).
Since observations of rich cluster abundance 
independently require that $b\approx 2$ for $\Omega_0=1$ 
(the cosmological mean density in units of the critical density)
inflationary models (Bahcall \& Cen 1992; \cite{wef93}; 
\cite{vl95};
\cite{ecf96}),
$b>2$ is necessary for the corresponding non-Gaussian (positively skewed)
models in order not to overproduce the abundance of 
rich clusters of galaxies.
We therefore conclude that
the texture model, 
or any random (or weakly correlated) seed cosmological models 
{\it per se}, is ruled out at a very high confidence level
($>3\sigma$).
For a related, very comprehensive study of non-Gaussian as well Gaussian
models using a suite of well defined statistical measures,
see Weinberg \& Cole (1992).

This paper is organized as follows.
In the next section we give an analytic technique to
compute the evolution of the correlation function
for any set of objects and apply it to the specific case
of clusters of galaxies in the texture model.
Conclusions are given in \S 3.

\section{Cluster Correlation Computed by an Analytic Technique}

The approach adopted here to calculate the cluster-cluster two-point
correlation function in the texture model is conceptually very simple,
as follows.
Barring merging, the 
temporal evolution of the two-point correlation
function of any set of objects, $\xi$, is governed by the 
pair conservation equation (equation 71.6 of Peebles 1980):
\begin{equation}
{\partial \xi\over\partial t}+{1\over x^2 a}{\partial\over\partial x}\left[x^2(1+\xi) v\right]=0 \quad ,
\end{equation}
\noindent 
where
$t$ is time; $a$ is the expansion parameter;
$x$ is the comoving separation;
and $v$ is the 
mean pairwise proper peculiar velocity.
Our subsequent results are directly or indirectly based on this equation.
Note that equation (1) is valid for {\it any set of objects}.
We will first derive a relation between 
pairwise peculiar velocity and correlation function 
of the matter in the linear regime,
using equation (1).
Following Subramanian \& Padmanahban (1994),
substituting the conventional two-point
matter correlation function $\xi_m$ with $\bar\xi_m$, 
the volume-averaged two-point
correlation 
of the {\it underlying matter}
within a sphere of radius $x$, defined by
\begin{equation}
\bar\xi_m (x,t) \equiv {3\over x^3}\int_0^x \xi_m(y,t) y^2 dy \quad , 
\end{equation}
\noindent
we may obtain a slightly different form
for equation (1) as:
\begin{equation}
\left({\partial\over\partial\ln a}-h{\partial\over\partial\ln x}\right)(1+\bar\xi_m)=3h(1+\bar\xi_m) \quad , 
\end{equation}
\noindent where $h\equiv v/(-\dot a x)$.
Note that $v$ is negative when pairs approach.
Equation (3) is obtained by substituting equation (2) into equation (1)
and by integrating it once with respect to $x$.
An integration constant on the right hand side of equation (3)
has been set to zero in order to 
produce the right asymptotic behavior:
$\xi_m$ freezes out in comoving space when $h=0$.
In the linear limit, where the peculiar velocity
is much smaller than the Hubble velocity (i.e., $h<<1$)
and $\bar\xi_m\propto (1+z)^{-2}$ (for an $\Omega_0=1$ universe, 
which is assumed throughout 
and consistent with the texture model where
$\Omega_0=1$ is usually assumed; we note that
this linear growth rate of matter correlation
is in agreement with detailed numerical simulations
of Cen \etal 1991 and  Park, Spergel, \& Turok 1991),
the only nontrivial solution is 
\begin{equation}
v(x,t) = {2\over 3} \bar\xi_m v_H \quad , 
\end{equation}
\noindent
where $v_H=\dot a x$ is the Hubble velocity 
for separation $x$ at time $t$.

Equation (4) allows us to obtain the evolution
of the total matter pairwise motion, $v(x,t)$,
from $\bar\xi_m$, which in turn is
constrained by the observed galaxy-galaxy two-point
correlation function.
For the linear scales which we are interested in, namely $\ge 10h^{-1}$Mpc,
where density fluctuations at all epochs
are in the linear regime,
we assume the following form:
\begin{equation}
\xi_m(x,z) = \left({x\over x_{gg,0}}\right)^{-\gamma} {1\over (1+z)^2} {1\over b^2} \quad ,
\end{equation}
\noindent where $x_{gg,0}$ is the galaxy-galaxy correlation length
at $z=0$; 
$b\equiv {\sigma_{gal,rms}\over \sigma_{m,rms}}$ is the (linear) bias factor;
and we have changed the time variable from $t$
to redshift $z$.
Combining equations (2,4,5) we obtain the 
{\it mean comoving} pairwise velocity of total matter
(similar to equation 71.13 of Peebles 1980),
$v_{mc} \equiv v(1+z)$:
\begin{equation}
v_{mc}(x,z) = {2\over 3-\gamma} {H_0 x_{gg,0}\over b^2}\left({x\over x_{gg,0}}\right)^{-\gamma +1} (1+z)^{-1/2} \quad , 
\end{equation}
\noindent 
where $H_0$ is the present Hubble constant
and we have made use of the relation $H(z)=H_0(1+z)^{3/2}$ for 
the $\Omega_0=1$ cosmology.

Equation (6) describes the evolution of the {\it mean comoving}
pairwise velocity of {\it total matter}
in the {\it linear regime}, constrained 
by the observed galaxy-galaxy correlation function and linear theory.
We now make a critical observation that
a subset of objects, for example, cluster-scale textures, 
can have {\it different} $v_{mc}$ from that of total matter.
It would not be hard to imagine cases where such situations
could arise.
For instance,
in defect models where density fluctuations
are produced by initial density and/or velocity kicks,
both decaying and growing modes exist,
whereas equation (6)
describes only the growing mode of the overall matter perturbation
under the self-gravitational action of the matter itself.
While detailed dynamics of textures and 
the induced matter perturbations are very complicated 
and require nonlinear numerical calculations,
we argue that there is a maximal $v_{mc}$
for any random set of objects
within the gravitational instability framework.
Two examples should demonstrate 
the relevant possible cases.
In the first example, if the perturbation seeds are densely populated
in redshift, one may obtain the time-averaged peculiar
velocity at any epoch as
\begin{equation}
v_{mc,dense}(z_f)\equiv {\int_{t_f}^\infty v_{mc}(z_f) (1+z)/(1+z_f) dt\over\int_{z_f}^\infty dt}=3v_{mc}(z_f) \quad ,
\end{equation}
\noindent 
where the second equality is obtained by inserting 
$t\propto (1+z)^{-3/2}$ and $v_{mc}$ is given by equation (6).
In the second example,
if the perturbation seeds are sparsely created
in redshift, one then has the time-averaged peculiar
velocity at any epoch as
\begin{equation}
v_{mc,sparse}(z_f)\equiv {\int_{t_f}^\infty v_{mc}(z_f) (1+z)/(1+z_f) dt\over\int_{z_f}^\infty (1+z_f)^{3/2}/(1+z)^{3/2} dt} =6v_{mc}(z_f) \quad .
\end{equation}
\noindent
The denominator in equation (8)
becomes apparent by noting that $v_{mc}\propto (1+z)^{-3/2}$ 
($v$ is the proper 
pairwise velocity),
indicated by equations (4,5).
In both cases (equations 7,8),
we have maximized the velocity
by assuming that the initial velocity (kick) is just let decay 
as $\propto (1+z)$ due to the universal expansion.
One may think of $v_{mc,dense}$ 
or $v_{mc,sparse}$ as 
the envelope wrapping a set of decaying velocity modes in redshift,
which are seamed onto the general matter velocity evolution 
(growing mode) when the decaying velocity is just about to
fall below the growing velocity.
Such decaying modes can be created, for example,
by velocity kicks due to
the unwinding of ever larger (when time progresses) 
textures in the 
expanding universe, up to the present (Spergel \etal 1991).

A critical assumption has just been made here and 
is worth being elaborated more.
That is, the velocity of these considered objects,
while allowed to be different from that of the matter for
periods of time,
is demanded to be synchronized with that
of overall matter at some intermittent  points.
This is equivalent to saying that, 
while perturbation seeds 
(whose occurrences may not be limited to certain time intervals)
create the initial perturbations,
the subsequent evolution of perturbed matter density/velocity fields
is determined by the gravitational instability of 
the matter itself (i.e., self-gravity of the matter).
However, if perturbation seeds not only initiate the perturbations 
but also dominate the subsequent evolution,
our assumption breaks down.
But such extreme
cases would not fit into the gravitational
instability picture which we are considering here;
i.e., such cases are likely to have some other grossly different 
properties than those encountered in the 
gravitational instability framework.
In any case, it seems that all proposed seed models, including those
based on topological defects or other, (perhaps)
more conventional seeds such as 
primordial black holes (e.g., Villumsen, Scherrer, \& Bertschinger 1991),
do not have such dominant perturbation seeds,
at least on the scales which we are investigating here
($x\ge 10h^{-1}$Mpc).
We therefore argue that
the pairwise velocity given in equation (8) 
constitutes the upper limit of possible values for any
set of selected objects
within the general gravitational instability framework.
In the specific case of the texture model,
the textures are relatively rare, as indicated
by the texture evolution results (Spergel \etal 1991),
$dn/d\eta = \nu/\eta^4$ with $\nu\sim 0.02$
and $n$ being the number density of textures created per unit comoving
volume ($\eta$ is the conformal time).
This implies that the correlation function of
clusters in the texture model
should be more appropriately upper limited 
by that derived using the velocity field given by equation (8),
as will be shown below.

We note that, while equation (1) is valid for
any set of objects,
Equation (4) may not hold for them even at large scales where
$v$ is small compared to the Hubble velocity (i.e., $h<<1$).
The reason is that equation (4) was derived in the limit 
where both $\bar\xi << 1$ and $h<<1$.
But the correlation of clusters of galaxies under consideration
here is supposedly significantly higher than that of matter
thus may not be much less unity.
As a result, we can not directly convert $v$ into $\xi$ using equation (4)
for the objects under our consideration,
rather we need to solve equation (1) directly by 
providing $v(x,t)$.
Although it is straightforward to solve equation (1) to
get $\xi$ given $v$, 
we find the following method (basically 
a Lagrangian integral form
of equation 1) conceptually simple
to work with to derive the evolution of $\xi$.
Let us write down the usual form (definition) of
$\xi$ at any epoch:
\begin{equation}
\xi(x)= {N_p\over 4\pi x^2 \Delta x n} - 1 \quad ,
\end{equation}
\noindent
where $N_p$ is the number of pairs
within a shell of width $\Delta x$ located $x$ from an object in question;
$n$ is the number density of the objects of interest.
In the absence of shell crossing 
of pairs, which holds for the form of $v_{mc}$ as
given in equation (6),
we can relate the final correlation
to the initial correlaton by
\begin{equation}
\xi_f(x_f)= {x_i^2\over x_f^2}{dx_i\over dx_f}[\xi_i(x_i)+1] -1  \quad ,
\end{equation}
\noindent 
where $x_i$ and $x_f$ are the initial and final
separations (bins) of pairs under consideration, respectively.
Therefore, all that we need to do
is to solve the evolution of $x$, which is very simple to follow.
What is yet left to be specified 
is the initial correlation function of textures,
$\xi_i(x)$.
Spergel \etal (1991) have shown that textures 
are initially uncorrelated (randomly distributed) on
scales larger than their horizon size.
For textures which we are interested in here,
their horizon size is about $10h^{-1}$Mpc, giving
a mass of $\sim 10^{15}h^{-1}M_\odot$.
Therefore, there is no clustering of cluster-size textures
on scales $\ge 10h^{-1}$Mpc {\it initially},
the range over which we wish to compute
the evolution of clustering of these textures;
i.e., $\xi_i(x)=0$ at $x>10h^{-1}$Mpc.
A related point, the majority of these textures are produced
at times slightly before the epoch of matter radiation equality at
redshift $z\sim 10^4$ in the texture model
(Gooding, Spergel, \& Turok 1991).

Integrating ${dx\over dt} = v_{mc}$, where $v_{mc}$ is given by
Equation (7) or (8), with
the initial condition $\xi_i(x)=0$
at $z_i$ (the starting redshift of pairwise movement,
presumably the birth redshift of the relevant textures at $z_i\sim 10^4$)
give the correlation function of the clusters at redshift $z$:
\begin{equation}
\xi(x,z) = (1+{y\over x^\gamma})^{{3\over \gamma}-1} -1 \quad , 
\end{equation}
\noindent
where 
\begin{equation}
y={\alpha\gamma\over 3-\gamma} x_{gg,0}^\gamma {1\over b^2} \left[{1\over (1+z)^2} - {1\over (1+z_i)^2}\right].
\end{equation}
\noindent
Setting $\xi$ to be unity (and $z_i=\infty$)
we can find the cluster-cluster correlation length at $z=0$
\begin{equation}
r_{cc,0} = x_{gg,0}\left[{\alpha\gamma \over (3-\gamma)(2^{\gamma \over 3-\gamma}-1)}\right]^{1/\gamma} b^{-2/\gamma} \quad , 
\end{equation}
\noindent
where $\alpha$ is $3$ and $6$, respectively,
for the two cases indicated by equations (7) and (8).
The results shown 
in Figure 1, the corrlation length
of clusters as a function of bias parameter $b$,
are computed using equation (13) 
with the canonical observed values for galaxies (Davis \& Geller 1976) 
of $\gamma=1.8$ and $r_{gg,0}=5.0h^{-1}$Mpc.
Three cases are shown for three values of $\alpha$:
$1$ corresponds to matter correlation (heavy solid curve),
$3$ corresponds to the upper limit in the case of densely populated seeds
in redshift space (heavy long dashed curve),
and $6$ corresponds to the upper limit in the case of 
sparsely populated seeds in redshift space (heavy short dashed curve),
which we argue is the absolute upper limit.
Also shown in Figure 1 are
the corresponding cases with $\gamma=1.0$ to illustrate
the dependence of the correlation length on $\gamma$,
the slope of the correlation function.
We note that, in the derivation of $r_{cc,0}$,
no assumption about the cluster number density is made,
except the very weak dependence of the calculation on $z_i$ 
due to different birth redshift of seeds for
 different mass clusters.
This implies that the cluster correlation length in
the texture model,
or any other seed model of that sort,
is richness independent.

We see that the cluster-cluster
two-point correlation length 
should be in the range 
$r_{cc,0}=5.0-13.0h^{-1}$Mpc (between heavy solid and short dashed curves)
for an unbiased $b=1$ model, and decreases
with increasing bias parameter.
This correlation length range for the $b=1$ model
is consistent with the results obtained
from N-body simulations by Park, Spergel, \& Turok (1991),
who find a correlation length of $\sim 9.0\pm 3.0h^{-1}$Mpc 
(adapted from Figure 3 of Park \etal 1991)
for the $b=1$ model.
This agreement lends
support for the validity of this analytic approach,
and perhaps explains the origin of cluster-cluster 
correlation in the texture model.
Another way to explain the origin is this.
The correlation of clusters of galaxies, whose seeds
are produced without any mutual correlation initially 
at high redshift near matter radiation equality
in the textures model,
grows gradually with time due to 
density perturbations on larger scales induced 
by larger textures which unwind at subsequent times 
up to the present.

Observations of galaxy cluster abundance
indicate that the bias parameter $b\sim 2$
for $\Omega_0=1$ inflationary Gaussian models
(Bahcall \& Cen 1992; \cite{wef93}; \cite{vl95}; \cite{ecf96}).
The required bias parameter in non-Gaussian (positively skewed)
models is even larger due precisely to 
the non-Gaussianity, which enables collapse of overdense regions easier.
We hence see that,
for a viable defect model with $b>2$ and $\Omega_0=1$,
the cluster-cluster correlation function length
should be smaller than
$r_{cc,0}=6.0h^{-1}$Mpc (obtained for $b=2$ models).
The depedence on $\gamma$ is weak at $b\ge 2$ as seen in Figure 1:
for the extreme case with $\gamma=1.0$, $r_{cc,0}=2.0-9.0h^{-1}$Mpc
for $b=2$ models.
The range in $r_{cc,0}$ is due to the possible range in $\alpha$.

Postman, Huchra, \& Geller (1992)
give $r_{cc,0}=20.0\pm 4.3h^{-1}$Mpc using
a complete sample of 361 Abell clusters.
Nichol \etal (1992) present $r_{cc,0}=16.4\pm 4.0h^{-1}$Mpc for
a 90 percent complete sample of 97 clusters from 
the Edinburgh-Durham Southern Galaxy Catalogue.
Dalton \etal (1992) yield $r_{cc,0}=12.9\pm 1.4h^{-1}$Mpc 
for 173 clusters from APM galaxy survey and 
$r_{cc,0}=14.0\pm 4.0h^{-1}$Mpc for the 93 richest clusters in the sample.
Dalton \etal (1994) conclude that 
$r_{cc,0}=14.3\pm 2.4h^{-1}$Mpc ($2\sigma$)
for 364 clusters from an extended APM galaxy survey.
Romer \etal (1994) indicate $r_{cc,0}=13-15h^{-1}$Mpc for
a ROSAT X-ray selected cluster sample of 128 clusters.
Croft \etal (1997) find $r_{cc,0}=21.3^{+11.1}_{-9.3}h^{-1}$Mpc ($2\sigma$)
for clusters with a mean space density of $1.6\times 10^{-6}h^{3}$Mpc$^{-3}$,
equivalent to the space density of Abell richness $\ge 2$ clusters,
by analysing a new catalogue of very rich clusters selected
from APM galaxy survey.
While uncertainties remain in the 
current clustering analyses of observed clusters
due to still limited cluster samples of order of a 
few hundred clusters at most,
it seems that 
all these studies have 
consistently given a correlation length 
$\ge 10h^{-1}$Mpc at approximately $2\sigma$ level with
average errorbar of size of $2-4\;h^{-1}$Mpc.
We therefore conclude that
$b\ge 2$ texture models 
can be ruled out at a very high
confidence level ($>3\sigma$).

\section{Conclusions}

Based on simple analytic reasoning within the gravitational
instability framework, we are able to set an upper limit 
on the correlation length of clusters of galaxies
in any {\it random} seed cosmological model, 
regardless of the nature of the seeds, which
could be topological defects such as textures or more
conventional seeds such as primordial black holes.
It is shown that clusters of galaxies
in any biased $b=2$ model has
a two-point correlation length of $r_{cc,0}\le 6.0h^{-1}$Mpc 
(for a slope of the correlation function of $-1.8$).
More likely, one is forced to adopt a bias parameter
$b<2$ for such non-Gaussian (positively skewed) models
as not to overproduce 
the abundance of rich clusters of galaxies observed locally,
since it is easier to form overdense structures 
with non-Gaussian positively skewed density perturbations.
All recent observations indicate 
a correlation length $r_{cc,0}>10.0h^{-1}$Mpc at an approximately
$2\sigma$ level for real rich clusters of galaxies
(with a typical errorbar size of $2-4\;h^{-1}$Mpc).
It is thus apparent that any topological defect models
or any random seed models 
are ruled out at a very high confidence level ($>3\sigma$).

While this constraint is completely independent of 
observations on any other scales including those of the CMB,
the requirement of a large $b\sim 4$ 
in order to fit the observed angular power of CMB 
on very large scales for the global 
texture model (Pen \etal 1997), for example, 
would render the model inviable in terms of matching
the observed correlation of rich galaxy clusters.

An unbiased $b=1$ topological defect model might
be able to provide a reasonable match to the
observed cluster correlation.
One way to achieve this is to 
adopt a lower density model with
$\Omega_0=0.2-0.3$ (with or without a cosmological
constant $\Lambda$).
But such a model might 
have more difficulty in fitting 
the large scale CMB observations.
In addition, such a model (especially in the case with zero $\Lambda$)
might run into the opposite of a problem 
that inflationary cold dark matter models have:
too early structure formation.
This deserves more careful calculations,
should such a model be put forth.

Another exit is to 
have the perturbations seeds responsible
for the formation of clusters of galaxies at later times
significantly clustered at birth before any gravitational evolution,
somewhat analagous to high density peaks in 
Gaussian models.
This might be an attractive route to search for potentially
viable seed models, especially those based on topological defects.

\acknowledgments
I would like to thank Michael Strauss for a careful
reading of the paper and the referee for a nice report.
Discussions with Neta Bahcall, Jerry Ostriker and 
David Spergel are also acknowledged.
The work is supported in part by grants NAG5-2759 and ASC93-18185.

\newpage

\figcaption[Figure 1]{
Cluster-cluster two-point correlation length 
as a function of bias factor $b$,
using equation (13) 
with the canonical observed values for galaxies (Davis \& Geller 1976) 
of $\gamma=1.8$ and $r_{gg,0}=5.0$.
Three cases are shown for three values of $\alpha$:
$1$ corresponds to matter correlation (heavy solid curve),
$3$ corresponds to the upper limit in the case of densely populated seeds
in redshift space (heavy long dashed curve),
and $6$ corresponds to the upper limit in the case of 
sparsely populated seeds in redshift space (heavy short dashed curve).
Also shown in Figure 1 are
the corresponding cases with $\gamma=1.0$ to illustrate
the dependence of the correlation length on $\gamma$,
the slope of the correlation function.
\label{fig1}}

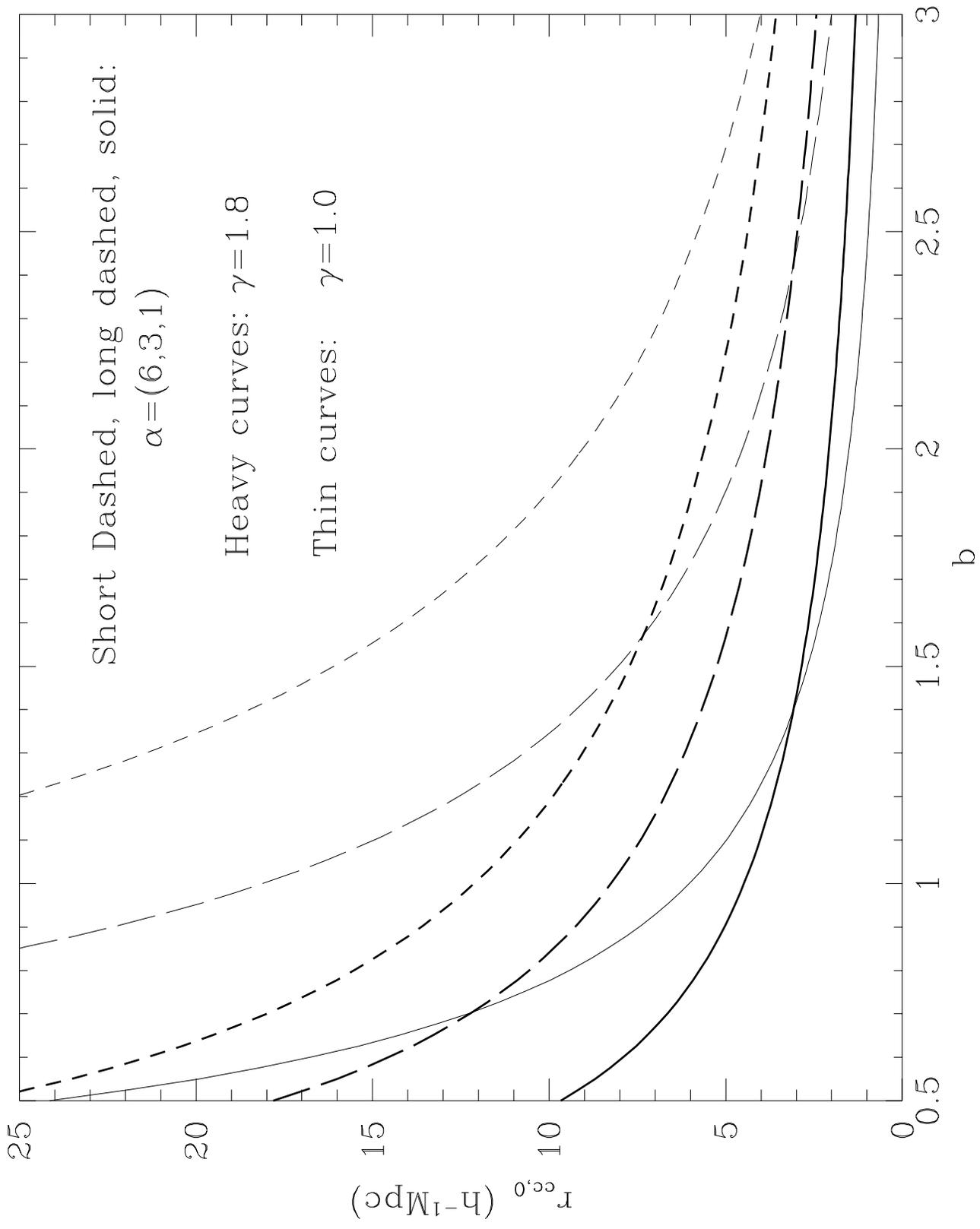
\end{document}